# Lightweight Weighted Average Ensemble Model for Pneumonia Detection in Chest X-Ray Images

**Suresh Babu Nettur[1*], Shanthi Karpurapu[1*], Unnati Nettur[2], Likhit Sagar Gajja[3], Sravanthy Myneni[1], Akhil Dusi[4] AND Lalithya Posham[5]**

[1] Independent Researcher, Virginia Beach, VA 23456, USA
[3] Department of Computer Science, Virginia Tech, Blacksburg, VA 24061 USA
[3] Department of Computer Science, BML Munjal University, Haryana 122413 INDIA
[4] Department of Information Systems, University of Indiana Tech, Indiana USA
[5] Nanjing Medical University, Nanjing City, Jiyangsu, CHINA

Corresponding authors: Shanthi Karpurapu (shanthi.karpurapu@gmail.com), Suresh Babu Nettur (nettursuresh@gmail.com)

[*] Shanthi Karpurapu and Suresh Babu Nettur are co-first authors

**ABSTRACT** Pneumonia is a leading cause of illness and death in children, underscoring the need for early and accurate detection. In this study, we propose a novel lightweight ensemble model for detecting pneumonia in children using chest X-ray images. Our proposed ensemble model integrates two pre-trained convolutional neural networks (CNNs), MobileNetV2 and NASNetMobile, selected for their balance of computational efficiency and accuracy. These models were fine-tuned on a pediatric chest X-ray dataset and combined to enhance classification performance. Our proposed ensemble model achieved a classification accuracy of 98.63%, significantly outperforming individual models such as MobileNetV2 (97.10%) and NASNetMobile(96.25%) in terms of accuracy, precision, recall, and F1 score. Moreover, the ensemble model outperformed state-of-the-art architectures, including ResNet50, InceptionV3, and DenseNet201, while maintaining computational efficiency. The proposed lightweight ensemble model presents a highly effective and resource-efficient solution for pneumonia detection, making it particularly suitable for deployment in resource-constrained settings.

## I. INTRODUCTION

Pneumonia remains a critical global health concern, accounting for 14% of all deaths of children under five years old and claiming the lives of approximately 740,180 children in 2019 [1]. This burden is disproportionately high in developing nations, where medical resources are scarce, and energy poverty exacerbates health risks. In 2020, household air pollution is responsible for approximately 3.2 million premature deaths annually. Among these, 21% are due to lower respiratory infections (LRIs). Exposure to household air pollution nearly doubles the risk of childhood LRIs and is responsible for 44% of all pneumonia deaths in children under five years old [2]. In such regions, the challenge is further compounded by limited access to healthcare infrastructure and severe shortages of medical personnel. For instance, there are 57 countries that face a shortfall of 2.3 million doctors and nurses [3] [4].

The accurate and timely diagnosis of pneumonia is essential to improving outcomes, especially in resource-constrained areas where early detection can save lives and reduce treatment costs. Radiological diagnostic techniques such as chest X-rays, magnetic resonance imaging (MRI), and computed tomography (CT) are widely used for lung diseases. Among these, chest X-rays are the most common due to their non-invasive nature and cost-effectiveness. However, the diagnosis of pneumonia using chest X-rays heavily relies on the expertise of radiologists, and the diagnostic process remains vulnerable to subjectivity and inconsistency, especially in regions with high disease burden [5] [6] [7].

To address these challenges, the development of automated diagnostic systems leveraging deep learning has emerged as a promising solution. Convolutional Neural Networks (CNNs), one of the most effective deep learning models, have demonstrated significant potential in medical image identification tasks by automatically extracting relevant features from images using backpropagation algorithms [8]. The CNN-based approaches not only minimize reliance on manual resources but also have the potential to promote efficiency, scalability, and reliability in pneumonia diagnosis. By integrating AI-based automated classification systems, radiologists' efficiency can be significantly enhanced. These AI advancements can be particularly beneficial for underserved regions by optimizing medical resource allocation and accelerating pneumonia diagnosis and treatment in children [9].

The central focus of our research is to enhance the application of deep learning and LLMs in software engineering [10] [11] and medical diagnostics [12][13]. Based on recent advancements in pneumonia detection, our goal is to develop a high-performing classification system that addresses the computational inefficiencies and limitations of existing CNN-based methods. To achieve this, we propose a novel weighted average ensemble approach using lightweight models designed to improve the detection of pneumonia from chest X-ray images. For this study, we utilized the Kermany dataset [14], a well-curated and widely recognized collection of chest X-ray images, as the primary data source for training and evaluation. Additionally, this dataset was used for comparative analysis to benchmark our proposed model against existing state-of-the-art methods, ensuring a comprehensive evaluation of performance and computational efficiency.

## II. RELATED WORKS
In recent studies, various approaches have been explored to improve pneumonia detection using deep learning techniques, with the Kermany dataset [14] [15], a collection of children's chest X-ray images, being commonly utilized in this research. In this section, we discuss the various research efforts conducted in this field, highlighting different methodologies and models applied to the Kermany dataset.

### A. PRE-TRAINED MODEL WITH TRANSFER LEARNING
Ayan et al. conducted a comparative study using this dataset, allocating 80% for training, 10% for validation, and 10% for testing [16]. They fine-tuned VGG16 and Xception models, achieving accuracies of 87% and 82%, respectively [16]. Similarly, Thakur et al. employed the Kermany dataset with 80% for training, 10% validation, and 10% testing and focused on fine-tuning VGG16, which resulted in an improved accuracy of 90.54% [17]. Jain et al. also used the Kermany dataset but with a 90% training and 10% testing split [18]. They developed six models, including two custom models and four pre-trained models (VGG16, VGG19, ResNet50, and InceptionV3). The custom models achieved validation accuracies of 85.26% and 92.31%, while the pre-trained models achieved accuracies of 87.28% (VGG16), 88.46% (VGG19), 77.56% (ResNet50), and 70.99% (InceptionV3) [18]. Chhikara et al. proposed a modified InceptionV3 model with pre-processing techniques such as gamma correction, JPEG compression, median filtering, and CLAHE [19]. Using the same dataset split (80% training, 10% validation, and 10% testing), they demonstrated the effectiveness of their approach for pneumonia diagnosis [19].

In contrast, Sara et al. employed InceptionV3 to extract features from chest X-ray images and trained three classification algorithms (K-Nearest Neighbor, Neural Network, and Support Vector Machines) on a Kaggle dataset [20]. Their Support Vector Machines model achieved the highest AUC score of 93.1% [20]. El et al. extended their study by combining the Kermany dataset (5,856 images) with an additional dataset of 231 COVID-related images [21]. They allocated 60% of the combined data for training and 40% for validation. Their work involved various pre-trained models, including InceptionV3, DenseNet201, VGG19, Xception, VGG16, InceptionResNetV2, MobileNetV2, a customized CNN, and ResNet50. Using data augmentation, MobileNetV2 and InceptionResNetV2 achieved accuracies of approximately 96% [21]. Knock et al. adopted a transfer learning approach using VGG16, splitting the dataset into 90% for training and 10% for testing. Their model achieved an accuracy of 94% [22].

### B. CUSTOM DEEP LEARNING TECHNIQUES
Liang et al. designed a novel network architecture with residual structures to learn the effective texture characteristics of lung tissue [23]. The network consists of 49 convolutional layers with ReLU activation, followed by a single global average pooling layer and two dense layers. Using 90% of the data for training and validation and 10% for testing, their model achieved an accuracy of 90.5% [23]. Stephen et al. proposed a CNN with four convolutional layers, using 63.5% of the data for training and 36.5% for validation, achieving an accuracy of 93.7% [24]. Raheel et al. developed an 18-layer Deep Convolutional Neural Network, allocating 89.7% of the data for training, 0.3% for validation, and 10% for testing [25]. Their model achieved an accuracy of 94.3%, with sensitivity and specificity of 99.0% and 88%, respectively [25]. Omar et al. utilized a CNN with five convolutional layers, splitting the data into 90% for training and 10% for testing, and achieved an accuracy of 87.65% [26].

Wu et al. employed image enhancement techniques alongside a CNN-RF model and GridSearchCV-based RF, using 66.7% of the data for training and 32.8% for testing [27]. Their model reported accuracy, precision, and specificity values of 97%, 90%, and 95%, respectively [27]. Rajaraman et al. developed a Deep Convolutional Neural Network combined with an atlas-based detection algorithm, using 90% of the data for training and 10% for testing [28]. Their model achieved an accuracy of 96.2% and a specificity of 85.9% [28]. Chakraborty et al. proposed a 17-layer network with three convolutional layers, achieving accuracy, recall, and precision values of 95.62%, 96%, and 95%, respectively [29]. Hasan et al. utilized a CNN with three convolutional layers, splitting the data into 80% for training and 20% for testing and validation, and achieved an accuracy of 96.24%, with precision, recall, and F1-score values of 94.19%, 91.82%, and 92.98%, respectively [30].

*C. ENSEMBLE TECHNIQUE*

Toğaçar et al. utilized existing CNN models such as AlexNet, VGG-16, and VGG-19 as feature extractors, specifically using the last fully connected layer of each model [31]. These extracted features were then fed into machine learning models, including Decision Trees (DT), k-nearest Neighbors (kNN), Linear Discriminant Analysis (LDA), Logistic Regression (LR), and Support Vector Machines (SVM). Feature selection was performed using the mRMR method, and the best results were achieved by combining all features from mRMR, with an accuracy of 99.41% using the Kermany dataset (70% training, 30% testing) [31]. Chouhan et al. proposed an ensemble model that combined the outputs from several pre-trained models, surpassing the performance of individual models [32]. This ensemble, which included Inception V3, ResNet, AlexNet, GoogleNet, and DenseNet121, achieved state-of-the-art results with an accuracy of 96.4% and a recall of 99.62% on the Kermany dataset, using a 90% training and 10% testing split [32].

Hashmi et al. introduced a novel weighted classifier approach that optimally combined predictions from several state-of-the-art deep learning models, including ResNet18, Xception, InceptionV3, DenseNet121, and MobileNetV3 [33]. The weighted classifier model achieved a test accuracy of 98.43% and an AUC score of 99.76% on the Kermany dataset, with 88.05% training and 11.95% testing [33]. El Asnaoui et al. focused on ensemble models fine-tuned from InceptionResNetV2, ResNet50, and MobileNetV2 [34]. In their experiments, they used two datasets: the first containing 5856 images from the Kermany dataset and the second containing only 231 images. They found that InceptionResNetV2 performed best as a single model, with an F1 score of 93.52%. However, an ensemble of ResNet50, MobileNetV2, and InceptionResNetV2 yielded superior performance, achieving an F1 score of 94.84%. Their study used an 80% training and 20% testing split, incorporating both chest X-ray and CT datasets, including a COVID chest X-ray dataset with 231 images [34].

Compared to existing ensemble approaches, our novel lightweight ensemble model, integrating MobileNetV2 and NASNetMobile, demonstrates promising results by significantly reducing computational resources and costs. The combination of MobileNetV2 and NASNetMobile for pneumonia classification is unique and has not been explored before, making this approach a novel contribution to the field. By leveraging these lightweight models together, we achieved high accuracy while maintaining computational efficiency, distinguishing our approach from traditional ensemble methods.

## III. METHODOLOGY

The proposed lightweight ensemble model for pneumonia detection is illustrated in Figure 1. The preprocessing of the pneumonia dataset is carried out first, ensuring the input images are ready for transfer learning. Three lightweight pre-trained CNN models, namely NASNetMobile, MobileNetV2, and EfficientNetB0, were selected for this study. These pre-trained models are enhanced by freezing initial layers and applying modifications such as GlobalAveragePooling2D, Dropout, and BatchNormalization, with Dense layers and a Sigmoid activation for classification. Next, these models are fine-tuned and evaluated on the dataset. Then, the top two performing models were selected to implement the weighted ensemble approach. The individual predictions, P1 and P2, from these models are then combined, with each prediction being assigned a weight (W1 and W2, respectively) to optimize the overall accuracy of the ensemble. Finally, the weighted ensemble model is developed using the optimal weights and comprehensively evaluated, effectively combining the strengths of the two selected CNN architectures for accurate pneumonia detection. In this section, we will discuss the key components of the methodology.

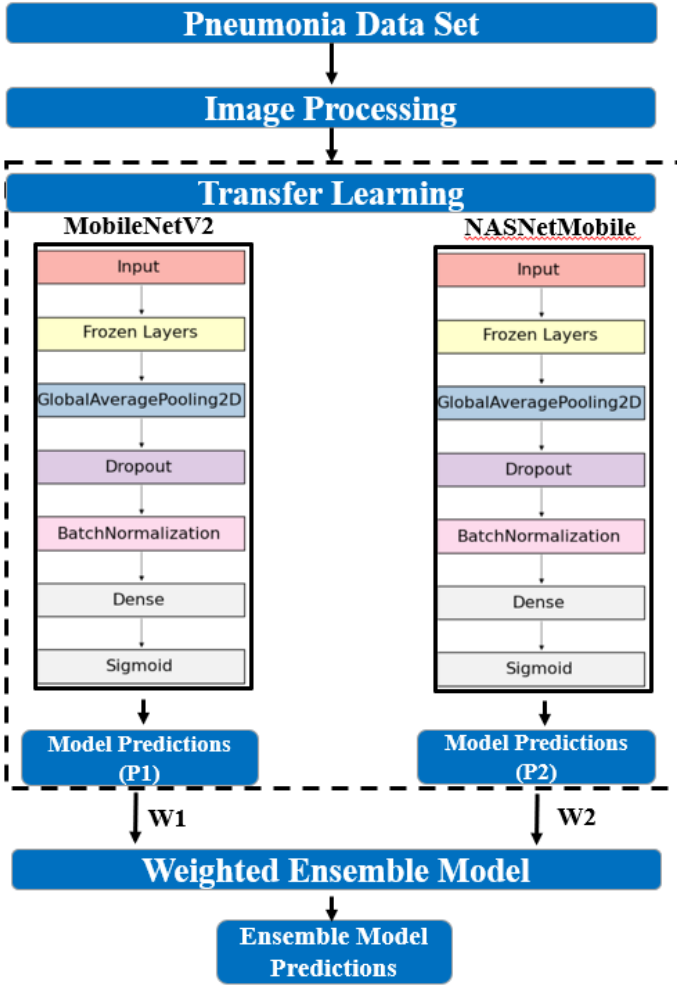

**FIGURE 1.** Weighted Average Ensemble Model Implementation

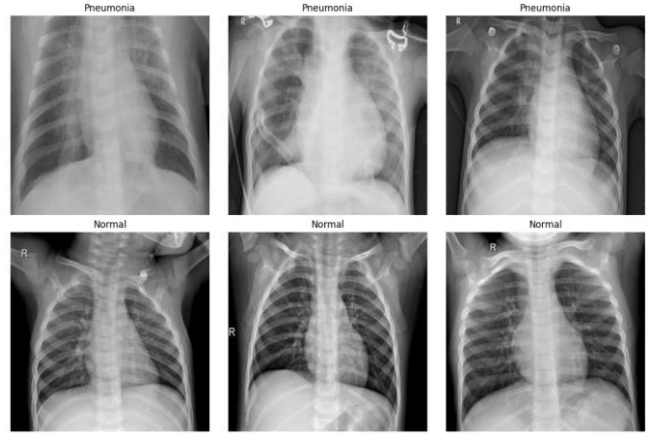

**FIGURE 2.** Sample Chest X-ray images of Normal and Pneumonia Patients

### A. DATASET
Kermany et al. developed the dataset by collecting chest X-ray images of patients under 5 years of age at Guangzhou Women and Children's Medical Center [14] [15]. The dataset comprises 5856 images with varying resolutions ranging from 400p to 2000p. It includes 4273 images of pneumonia cases and 1583 images of normal cases. The dataset is utilized for training, and the performance of various transfer learning models is evaluated. Some of these images are shown in Figure 2. For effective model training and evaluation, the dataset was divided into two parts: 90% of the images were allocated for training, while the remaining 10% were allocated for testing. As the next step, image pre-processing techniques and augmentation strategies were applied to enhance the quality and variability of the dataset, ensuring it was well-suited for effective model training and evaluation.

### B. IMAGE PRE-PROCESSING AND AUGMENTATION
In computer vision tasks, image pre-processing plays a vital role in preparing data for model training. Pre-processing techniques enhance model performance by addressing key factors such as noise reduction, image normalization, and resizing. These steps are especially critical for CNNs, which depend on consistent input dimensions and appropriately scaled pixel values. In our work, pixel intensity normalization was applied to scale the pixel values to the range [0, 1]. This step is essential for stabilizing the training process and facilitating efficient model convergence, allowing the model to learn patterns effectively without being affected by variations in image brightness or contrast.

Furthermore, the images were resized to ensure compatibility with the input dimensions required by the network. This resizing step is crucial for maintaining consistency across the dataset, particularly when utilizing pre-trained models that mandate fixed input sizes. The choice of 224x224 pixels aligns with widely adopted practices in computer vision, as these dimensions are used in models such as MobileNetV2, NASNetMobile, and EfficientNetB0. This ensures not only compatibility but also efficient training and inference.

To enhance dataset diversity and simulate real-world variations, we applied image augmentation, a critical step for building a more resilient model and minimizing overfitting. Our approach involved implementing a range of transformations, including rotation, width and height shifts, shear, zoom, and brightness adjustments. For example, slight rotations and positional shifts enabled the model to identify features from varying perspectives, while brightness and zoom modifications allowed the model to generalize across different lighting conditions and scales. These augmentations ensure that the model learns more robust and adaptable features, improving its ability to manage a wide range of real-world scenarios and data variations. Table 1 outlines the augmentation techniques in detail, including the specific parameters applied for each transformation. The prepared and augmented dataset is essential for the next step of applying transfer learning to fine-tune pre-trained CNN models for pneumonia detection.

TABLE I
IMAGE AUGMENTATION PARAMETER SETTINGS

| Augmentation Parameter | Value | Description |
|---|---|---|
| Rotation Range | ±5 degrees | Randomly rotates images within ±5 degrees |
| Width Shift Range | 5% | Shifts images horizontally by up to 5% of the image width |
| Height Shift Range | 5% | Shifts images vertically by up to 5% of the image height |
| Shear Range | 0.05 | Applies a random shear transformation with intensity of 0.1 |
| Zoom Range | 5% | Randomly zooms in or out by up to 5% |
| Brightness Range | [0.9, 1.1] | Randomly adjusts brightness within the specified range |
| Fill Mode | Reflect | Fills points outside the boundaries by reflecting edges |

### C. TRANSFER LEARNING

We employed a transfer learning approach to fine-tune pre-trained CNN models, MobileNetV2, EfficientNetB0, and NASNetMobile, for Pneumonia detection. These models were originally trained on ImageNet, a large-scale dataset consisting of 1.28 million natural images across 1,000 categories. Utilizing the feature representations learned from extensive datasets like ImageNet reduces the reliance on large amounts of labeled data, which is often limited during pandemics. It also enabled quick adaptation to the specific task while improving diagnostic accuracy. Choosing transfer learning offers a scalable and cost-effective solution for enhancing Pneumonia detection, effectively addressing challenges related to data scarcity and limited resources. In this study, we applied transfer learning to specifically include lightweight CNN models, which are specifically designed to meet the unique needs of mobile and edge devices.

### D. TRANSFER LEARNING WITH LIGHTWEIGHT MODELS

Lightweight CNN models have a significantly reduced number of parameters, with smaller model sizes, faster detection speeds, and lower memory usage. They are particularly effective in medical image analysis, such as detecting lung pathologies, tumors, and heart conditions, where accurate and fast diagnoses are crucial [35] [36][37][38]. Their ability to operate efficiently in low-cost, real-time settings is vital for practical deployment in healthcare and other resource-limited environments, such as mobile phones or edge devices. The selection of MobileNetV2, EfficientNetB0, and NASNetMobile for our study was based on their strong performance in state-of-the-art reviews, with a focus on lightweight models that deliver efficient performance. The lightweight model details(https://keras.io/api/applications/) are shown in Table 2. Top-5 accuracy mentioned in Table 2 refers to the percentage of times the correct label is within the model's top 5 predicted labels. Next, we discuss the specifics of each model, highlighting their unique features and performance characteristics.

TABLE II
LIGHTWEIGHT DEEP LEARNING PRE-TRAINED MODEL DETAILS

| Model | Size (MB) | Top-5 Accuracy on ImageNet Dataset | Parameters | Time (ms) per inference step (CPU) |
|---|---|---|---|---|
| MobileNetV2 | 14 | 90.1% | 3.5M | 25.9 |
| NASNetMobile | 23 | 91.9% | 5.3M | 27 |
| EfficientNetB0 | 29 | 93.3% | 5.3M | 46 |

### E. MOBILENETV2

MobileNetV2, introduced by Sandler et al. (2018) [39], is a CNN architecture specifically designed for mobile and embedded vision applications. It employs an inverted residual structure with shortcut connections between compact bottleneck layers, which effectively reduces the number of parameters while enhancing computational efficiency. The architecture begins with a 32-filter convolutional layer, followed by 19 bottleneck layers that enable the construction of deeper networks without significantly increasing the size of intermediate layers. MobileNetV2 is particularly well-suited for tasks demanding efficient performance, such as image segmentation, object detection, and real-time inference on mobile and edge devices [40] [41] [42] [43] [44] [45].

### F. EFFICIENTNETB0

EfficientNet is a convolutional neural network (CNN) architecture that introduces compound scaling, which optimally balances the width, depth, and resolution of the network to improve performance while maintaining efficiency [46]. EfficientNet is particularly well-suited for tasks requiring a balance of high accuracy and computational efficiency, such as image classification, object detection, and transfer learning, making it ideal for deployment on both resource-constrained devices and high-performance systems [47] [48] [49] [50] [51] [52] [53] [54] [55]. Width scaling applies a feature map to each layer, depth scaling increases the number of layers in the network, and resolution scaling enhances the input image resolution [56]. The EfficientNet family comprises eight variants, ranging from EfficientNet-B0 to EfficientNet-B7. The architecture we chose for this study is EfficientNet-B0, which incorporates the (Mobile Inverted Bottleneck

Convolution) MBConv block, an inverted bottleneck residual block from MobileNetV2. The MBConv block is specifically designed to optimize network efficiency [39].

### G. NASNETMOBILE
The NAS (Neural Architecture Search) [57] framework, developed by Google, is a scalable CNN architecture designed using reinforcement learning to configure its fundamental building blocks. Each cell comprises a small set of operations, such as convolutions and pooling, which are replicated multiple times to meet the required network capacity. Its lightweight variant, NasNetMobile, features 12 cells containing 5.3 million parameters and 564 multiply-accumulate operations, making it highly efficient for mobile and resource-constrained environments. NasNetMobile is particularly well-suited for tasks requiring efficient performance, including image classification, object detection, and real-time inference on mobile and edge devices [58] [59] [60] [61] [62].

### H. FINE-TUNING PRE-TRAINED MODELS
After selecting lightweight pre-trained CNN architectures, we fine-tuned them for the Pneumonia detection task. By leveraging pre-trained CNN architectures, we retained several of their initial layers, which were frozen to preserve the general features learned from ImageNet. Freezing these layers ensured that their weights were not updated during training on the new dataset. This prevented the model from overwriting essential, general-purpose feature representations such as edges, textures, and basic shapes learned from the diverse ImageNet dataset. The final classification layer was replaced with new layers tailored specifically for Pneumonia detection. Only these newly added layers were left unfrozen, allowing their weights to be updated during training. This targeted fine-tuning strategy improved task-specific performance for Pneumonia detection while maintaining computational efficiency. The fine-tuning approach we implemented is depicted in Figure 3.

As illustrated in Figure 3, the custom layers we added include a global average pooling layer to summarize the learned features. This layer reduces the spatial dimensions of the feature maps, improving computational efficiency and helping prevent overfitting by generating compact feature representations [63]. A dropout layer is also incorporated to mitigate overfitting by randomly omitting a portion of the units during training. Dropout serves as a regularizer, forcing the network to rely on a subset of neurons, which has been shown to enhance generalization by simulating a bagged ensemble of neural networks [64]. Additionally, a batch normalization layer is used to improve training stability and speed by normalizing the output of each layer, ensuring zero mean and unit variance [65]. This technique helps accelerate convergence and stabilize the learning process. To introduce non-linearity, the ReLU (Rectified Linear Unit) activation function is applied. ReLU, defined as f(x) = max (0, x), is widely used due to its simplicity, efficiency, and ability to mitigate the vanishing gradient problem while promoting sparsity. Finally, the dense layer employs a sigmoid activation function to output probabilities for the binary classification task. The sigmoid function, as defined in Equation 1, transforms inputs into a value between 0 and 1, effectively representing the probability of the positive class in a binary classification setting.

$$\sigma(x) = \frac{1}{(1+e^{-x})} \quad (1)$$

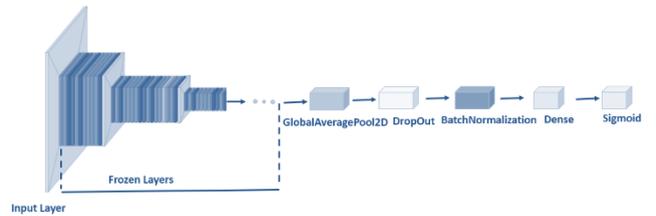

**FIGURE 3.** Transfer learning approach

We present a detailed comparison of the training parameters for three widely used architectures, EfficientNetB0, NASNetMobile, and MobileNetV2, under two distinct scenarios, as illustrated in Figure 4. The first scenario involves training all layers of the models from scratch, where we report the trainable parameters for each architecture. In contrast, the approach we followed is the second scenario, which focuses on training only the newly added layers through transfer learning. This approach significantly reduces the number of trainable parameters, as the pre-trained layers (frozen) retain useful learned features. Consequently, the reduced parameter count enables faster training with fewer computational resources.

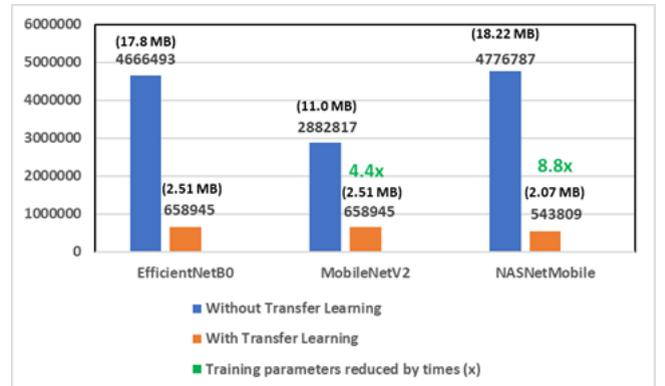

**FIGURE 4.** Comparison of trainable parameters for EfficientNetB0, MobileNetV2, and NASNetMobile With and Without Transfer Learning

## I. ENSEMBLE LEARNING

Ensemble learning has proven to be an effective approach for image classification, as it combines the predictions of multiple classifiers to achieve superior performance compared to individual classifiers. These models take advantage of the unique strengths of each classifier, enhancing accuracy and robustness by capturing different features of the data while compensating for their individual strengths and weaknesses. Bagging, boosting, and stacking methods are commonly used to build ensemble classifiers. Bagging, or Bootstrap Aggregating, creates diverse models by training each classifier on a bootstrapped subset of the training dataset. Boosting, on the other hand, iteratively improves weak learners by focusing more on misclassified examples, thereby boosting their performance over successive iterations. Stacking uses a meta-classifier that combines the predictions of base classifiers, utilizing these outputs as inputs to produce a higher-level representation of the data. By integrating multiple classifiers, ensemble learning provides a promising way to advance image classification, effectively improving performance. Ensemble methods are resistant to noise and overfitting, making them effective for complex, noisy image datasets. As a result, ensemble learning has become increasingly popular in various computer vision tasks, including medical image analysis and object recognition. However, it is important to note that ensemble learning may incur higher computational costs and resource demands due to the need for training and combining multiple models. To address these challenges, we selected lightweight models, which are more computationally efficient yet capable of yielding superior ensemble performance.

## J. WEIGHTED AVERAGE ENSEMBLE MODEL

In this study, we employed a weighted ensemble technique to enhance the classification performance for pneumonia. In order to implement this, we initiated a comprehensive model selection and training process. After training the three lightweight pre-trained deep learning models, namely EfficientNetB0, MobileNetV2, and NASNetMobile, their performance is evaluated. Based on individual model performance, we selected the top two models, MobileNetV2 and NASNetMobile, which demonstrated better prediction accuracies compared to the other models. Therefore, these two models were chosen as foundational models and combined to implement a weighted average ensemble (WAE) model. The WAE approach was chosen to leverage the strengths of individual models and to enhance the overall predictive capability. The weights for each base classifier's prediction were determined to maximize the WAE model accuracy.

The calculation of the WAE model predictions is shown in Equation 2. The WAE model prediction is represented as $y_{ensemble}$ and prediction of i$^{th}$ classifier is denoted as $y_i$, weight of i$^{th}$ classifier is denoted as $W_i$. The maximum accuracy is computed by evaluating the performance of the WAE model on test dataset. Each model's predictions are combined using a weighted average, where the weights are systematically varied to identify the optimal combination. Specifically, a grid search over all possible weight combinations is performed, with weights ranging from 0 to 1 in increments of 0.005, ensuring that the sum of weights equals one (Equation 3). The accuracy of the WAE model for each weight combination is calculated, and the weight combination yielding the highest accuracy represents the optimal weights for the ensemble, providing the maximum predictive performance.

$$\hat{y}_{ensemble} = \sum_{i=1}^{N} \omega_i \cdot \hat{y}_i \qquad (2)$$

$$\sum_{i=1}^{n} \omega_i = 1, \omega_i \in [0,1] \qquad (3)$$

### 1) Evaluation

We evaluated the performance of our proposed WAE model in comparison to the fine-tuned models, assessing their effectiveness in pneumonia detection using key metrics, including accuracy, recall, precision, F1 score, the area under the ROC curve, and the confusion matrix. These metrics provide a comprehensive evaluation of each model's ability to distinguish between pneumonia-positive and normal cases, offering insights into class-specific performance and overall classification effectiveness. This thorough evaluation enabled a clear comparison between the fine-tuned models and our proposed WAE model.

*Accuracy:*
Accuracy is an evaluation metric that measures the ratio of correct predictions (both true positives and true negatives) to the total number of predictions. It provides an overall assessment of each model's performance, as shown in Equation 4. Accuracy is calculated using the following components: TP (True Positives), TN (True Negatives), FP (False Positives), and FN (False Negatives).

$$Accuracy = \frac{TP+TN}{TP+TN+FP+FN} \qquad (4)$$

*Confusion Matrix:*
The confusion matrix provides a detailed breakdown of the model's performance for each class, displaying the counts of True Positives, True Negatives, False Positives, and False Negatives. This metric is essential for evaluating class-specific performance and identifying potential issues, such as class imbalances or misclassifications.

*Precision, Recall, and F1 Score:*

Precision, Recall, and F1 Score offer class-specific insights, particularly when evaluating model performance on imbalanced datasets. These metrics are derived from the confusion matrix. Precision (Equation 5) is the ratio of correctly predicted positive cases (True Positives) to all predicted positives, reflecting the accuracy of positive predictions. Recall (Equation 6) is the ratio of correctly predicted positives to all actual positives, indicating the model's ability to detect pneumonia cases. The F1 Score (Equation 7) is the harmonic mean of Precision and Recall, providing a balanced measure of performance in detecting pneumonia.

$$Precision = \frac{TP}{TP+FP} \qquad (5)$$

$$Recall = \frac{TP}{TP+FN} \qquad (6)$$

$$F1 = 2 \times \frac{Precision \times Recall}{Precision+Recall} \qquad (7)$$

*Weighted Average:*
In classification performance metrics, weighted averages are often used to aggregate Precision, Recall, and F1 scores across multiple classes, particularly in imbalanced datasets. These averages offer a more comprehensive view of model performance by accounting for both class imbalance and individual class performance, extending the interpretation of results beyond individual class metrics. The weighted average (Equation 8) is the mean of the metrics (Precision, Recall, F1 Score) for each class, weighted by the support or the number of instances for each class. This approach reflects the class distribution, providing a more realistic representation of model performance in imbalanced datasets. Larger classes have a greater influence on the weighted average, making it an ideal metric for evaluating overall model performance. The weighted average is calculated by summing the metric values $M_i$ for each class, where $M_i$ represents the metric for the $i$-th class, and multiplying each by the support $n_i$, the number of instances in the $i$-th class. The sum of these weighted values is then divided by the total number of instances $N$ in the dataset.

$$Weighted\ Average = \frac{1}{N} \sum_{i=1}^{c} M_i \times n_i \qquad (8)$$

*Area Under the Curve (AUC) and Receiver Operating Characteristic (ROC) Curve:*
The AUC score measures the model's ability to distinguish between pneumonia and normal cases. It is derived from the ROC curve, which plots the True Positive Rate (TPR) against the False Positive Rate (FPR) across different classification thresholds. An AUC score of 1.0 indicates perfect discrimination, while a score of 0.5 suggests random guessing. The AUC is calculated as shown in Equation 9. The ROC curve visualizes a binary classifier's diagnostic performance by plotting the TPR against the FPR as the discrimination threshold is varied. TPR and FPR are calculated as shown in Equations 10 and 11.

$$AUC = \int_0^1 TPR(t)\, d(FPR(t)) \qquad (9)$$

$$TPR = \frac{TP}{TP+FN} \qquad (10)$$

$$FPR = \frac{FP}{FP+TN} \qquad (11)$$

## IV. RESULTS

We conducted the experiments using Google Colab, utilizing CPU resources, 51 GB of RAM, and 225.8 GB of disk space. Python 3 and relevant libraries, including Scikit-Learn, Keras, and TensorFlow, were used to implement the proposed weighted average ensemble model. The pre-trained lightweight models, specifically MobileNetV2, EfficientNetB0, and NASNetMobile, were loaded from Keras, each initialized with ImageNet weights, and the top two performing models were incorporated into the WAE model for enhanced performance. We trained the three pre-trained learning CNN models using 5,270 normal and pneumonia patient X-ray images.

For model compilation, we used the Adam optimizer with a learning rate of 1e-4, combined with a binary cross-entropy loss function, which is suitable for binary classification tasks. To improve the training process, we incorporated several callbacks. Early stopping was applied to prevent overfitting by monitoring the validation loss and restoring the best weights if no improvement was observed after five epochs. We also implemented a learning rate reduction strategy, which adjusts the learning rate by a factor of 0.5 when a plateau in validation loss is detected, with a minimum learning rate of 1e-6. Additionally, model checkpointing was used to save the best-performing model based on validation loss, ensuring the retention of the most effective model after training. The training process was carried out on the augmented data for 20 epochs with a batch size of 16, utilizing the specified callbacks to optimize both performance and training efficiency. We compared the performance of the fine-tuned CNN model with our proposed WAE using 586 images, consisting of 423 Pneumonia-infected and 163 non-infected images. The models were evaluated based on various metrics outlined in the methodology section.

Before implementing the ensemble model, we evaluated the three lightweight fine-tuned CNN models to select the top two models with the highest accuracy. The accuracies of the models are presented in Table 3.

TABLE III
COMPARISON OF MODEL ACCURACY

| Model | Accuracy |
|---|---|
| MobileNetV2 | 97.10% |
| NASNetMobile | 96.25% |
| EfficientNetB0 | 72.18% |

The WAE model was implemented using MobileNetV2 and NASNetMobile. By following the approach detailed in the methodology section to identify optimal weights for achieving maximum accuracy, we found that NASNetMobile and MobileNetV2, with weights of 0.55 and 0.45, yield the highest WAE model accuracy. We generated a confusion report for fine-tuned models and the WAE model, as shown in Table 4, to assess their robustness by calculating accuracy, precision, recall, and F1 score in Table 4. Class-level metrics, including confusion reports for pneumonia and normal cases, are presented in Table 5. The confusion matrices for all models are depicted in Figure 5, while the ROC curves for all models are shown in Figure 6.

TABLE IV
FINE-TUNED CNN MODELS AND WAE MODEL CLASSIFICATION REPORTS
(MOBILENETV2 WEIGHT =0.45 AND NASNETMOBILE WEIGHT =0.55)

| Model | Accuracy | Precision (Weighted Avg) | Recall (Weighted Avg) | F1 (Weighted Avg) |
|---|---|---|---|---|
| NASNetMobile | 96.25% | 92.22% | 94.48% | 93.33% |
| MobileNetV2 | 97.10% | 97.12% | 97.10% | 97.11% |
| Proposed WAE Model | 98.63% | 98.66% | 98.63% | 98.64% |

TABLE V
FINE-TUNED CNN MODELS AND WAE MODEL CLASS WISE CLASSIFICATION
REPORTS (MOBILENETV2 WEIGHT =0.45 AND NASNETMOBILE WEIGHT =0.55)

| Model | Class | Precision | Recall | F1 score |
|---|---|---|---|---|
| NASNetMobile | PNEUMONIA | 97.85% | 96.93% | 97.39% |
|  | NORMAL | 92.22% | 94.48% | 93.33% |
| MobileNetV2 | PNEUMONIA | 98.33% | 97.64% | 97.98% |
|  | NORMAL | 93.98% | 95.71% | 94.83% |
| Proposed WAE Model | PNEUMONIA | 99.52% | 98.58% | 99.05% |
|  | NORMAL | 96.41% | 98.77% | 97.58% |

Our proposed WAE model, utilizing NASNetMobile and MobileNetV2 with the optimum weights, demonstrates outstanding performance with an accuracy of 98.63%. The weighted average precision, recall, and F1-score of the WAE model reach 98.66%, 98.63%, and 98.64%, respectively. Our proposed WAE model shows notable improvements compared to the individual models, MobileNetV2 (97.10%) and NASNetMobile (96.25%). In terms of class-wise performance, our WAE model achieves an impressive 99.52% precision and 98.58% recall for the pneumonia class, with a remarkable F1-score of 99.05%. For the normal class, our WAE model achieves a precision of 96.41%, recall of 98.77%, and F1-score of 97.58%. This demonstrates the superior performance of our WAE model across both classes.

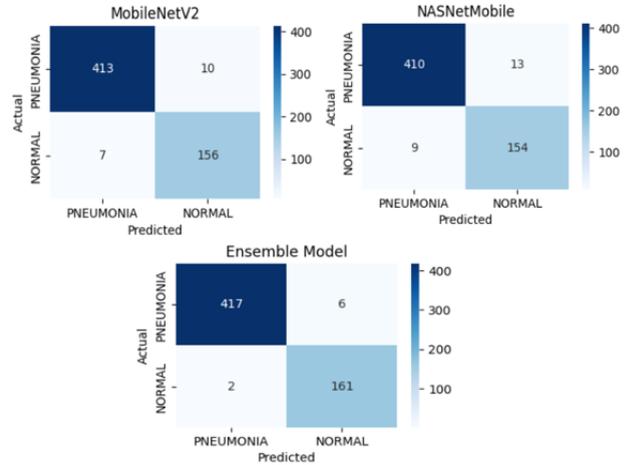

FIGURE 5. Confusion matrix for model evaluation

The confusion matrices for MobileNetV2, NASNetMobile, and the Ensemble Model shown in Figure 5 illustrate their respective performances in classifying pneumonia and normal cases. MobileNetV2 shows strong results, correctly identifying 413 pneumonia cases and 156 normal cases, though it misclassifies 10 pneumonia cases as normal and seven normal cases as pneumonia. Similarly, NASNetMobile performs well, accurately classifying 410 pneumonia cases and 154 normal cases, but it records slightly higher misclassifications, with 13 pneumonia cases predicted as normal and nine normal cases as pneumonia. In contrast, we observed the WAE Model demonstrates superior performance, correctly identifying 417 pneumonia cases and 161 normal cases while maintaining minimal errors, with only six pneumonia misclassifications and two normal misclassifications. These results highlight our proposed WAE Model's enhanced accuracy, recall, precision, F1, and overall performance compared to the individual models.

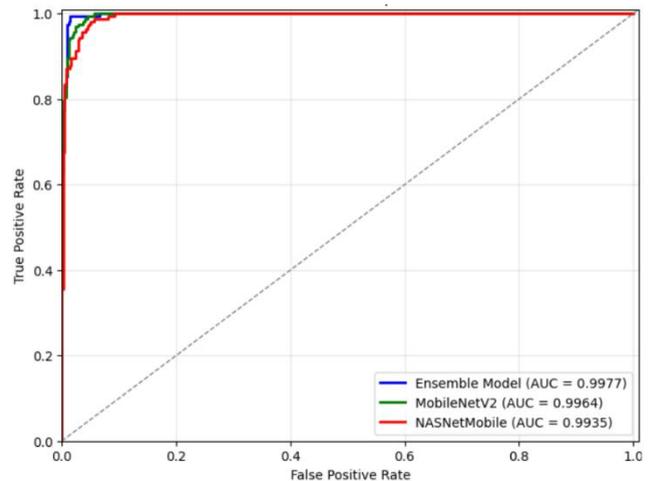

**FIGURE 6. ROC curve for different models**

The ROC curve shown in Figure 6 compares the performance of three models: WAE Model, MobileNetV2, and NASNetMobile. The True Positive Rate (TPR) is plotted against the False Positive Rate (FPR), demonstrating each model's ability to distinguish between classes. We observed that the WAE Model achieved the highest Area Under the Curve (AUC) score of 0.9977, indicating superior performance in classification. MobileNetV2 follows with an AUC of 0.9964, while NASNetMobile shows a slightly lower AUC of 0.9935.

Our proposed WAE model achieves an impressive accuracy of 98.63%, as shown in Figure 7, outperforming state-of-the-art and commonly used deep learning architectures. Compared to individual models like EfficientNetB2 (72.18%) and MobileNetV3Large (88.23%), the ensemble model demonstrates a substantial improvement in accuracy. Furthermore, it outperforms more complex architectures, including ResNet50 (93.34%), InceptionV3 (94.71%), and DenseNet201 (97.78%).

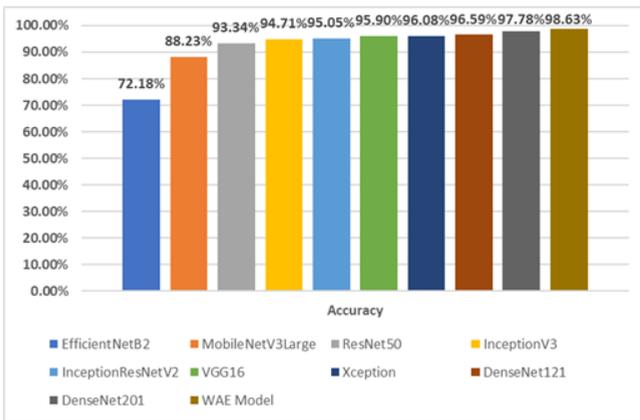

**FIGURE 7.** Accuracy Comparison of the WAE Model with State-of-the-Art and Common Deep Learning Architectures

## V. Discussions

The superior performance of our proposed weighted average ensemble (WAE) model, combining MobileNetV2 and NASNetMobile, can be attributed to the complementary strengths of these architectures and the benefits of the ensemble approach. MobileNetV2, with its lightweight architecture, achieves efficient feature extraction using depthwise separable convolutions and inverted residuals. This design enables MobileNetV2 to capture fine-grained features while maintaining low computational overhead, making it particularly effective for scenarios requiring precision and efficiency. Conversely, NASNetMobile, developed through Neural Architecture Search (NAS), specializes in identifying task-specific optimal architectures, enabling it to analyze intricate patterns such as fine-grained anomalies and texture variations in medical imaging datasets.

As demonstrated by the results, the WAE approach leverages the strengths of these models through a weighted combination of predictions, optimizing accuracy and robustness. The use of weighted averaging, with weights determined based on validation performance, ensures a balanced contribution from each model, addressing their individual limitations. This ensemble is particularly valuable in medical imaging applications, where diverse feature extraction is critical for accurate classification. Our WAE approach reduces the risk of overfitting by combining the diverse learning patterns of MobileNetV2 and NASNetMobile, each excelling in distinct feature representations. By integrating these architectures, the WAE model avoids over-reliance on dataset-specific features, thereby enhancing its ability to generalize to unseen data. This improved robustness is reflected in the WAE model's superior evaluation metrics, underscoring its capability to provide accurate and reliable predictions.

One notable advantage of the WAE model, as demonstrated by our results, is its ability to address the challenges posed by class imbalance. The WAE model achieves substantial improvements in precision, recall, and F1-score for the underrepresented NORMAL class compared to individual models. This balanced performance across both classes highlights the ensemble's capability to leverage the complementary strengths of MobileNetV2 and NASNetMobile.

Despite the promising results, our study has limitations that warrant further exploration in future work. The model was trained and evaluated on a single dataset, the Kermany dataset, which was chosen for its high-quality, well-curated, and widely recognized collection of chest X-ray images. While the dataset is an excellent foundation for our current study, it may not fully capture the diversity of image conditions, demographic variations, and disease patterns seen in real-world clinical settings. Variations in imaging equipment, patient populations, and environmental factors could impact the model's performance when applied to different healthcare environments. In the future, we plan to prioritize evaluating the model's adaptability and robustness across larger and more diverse datasets to ensure broader applicability.

Furthermore, while our study focuses on binary classification (pneumonia vs. normal), extending the model to multi-class classification tasks, such as distinguishing between different types or severities of pneumonia, could significantly enhance its clinical relevance. Real-time testing and collaboration with healthcare professionals are also crucial to evaluate the model's practical utility and usability in diagnostic workflows. Although our study meets its primary objectives of improving pneumonia detection, future work should investigate the performance of the proposed weighted average ensemble approach on larger, more diverse datasets to assess its scalability and adaptability. Exploring alternative data augmentation strategies, advanced optimization techniques, and hyperparameter tuning could further enhance model robustness and accuracy. Additionally, like many deep learning models, the proposed approach functions as a black box, offering limited interpretability of its predictions. Incorporating explainable AI techniques in future work could increase transparency and trust in the model's decision-making, making it more suitable for clinical deployment.

## VI. CONCLUSION

In this study, we proposed a novel weighted average ensemble (WAE) model for pneumonia detection in X-ray images, which combines the complementary strengths of two lightweight pre-trained models, MobileNetV2 and NASNetMobile. By leveraging these models, we achieved significant performance improvements over individual models, with an accuracy of 98.63% and precision, recall, and F1-score of 98.66%, 98.63%, and 98.64%, respectively. Our proposed ensemble model outperformed several pre-trained complex architectures, demonstrating its potential for real-time, resource-efficient applications in medical imaging. We highlighted the ensemble approach's effectiveness in balancing model strengths and weaknesses, providing superior performance while maintaining lower computational requirements compared to more complex models. The weighted averaging mechanism further optimized prediction accuracy, enhancing the overall performance of the model. In future work, we plan to explore advanced optimization techniques, expand the model's applicability to larger and more diverse datasets, and investigate its potential across different healthcare domains.

**REFERENCES**

[1] https://www.who.int/news-room/fact-sheets/detail/pneumonia
[2] https://www.who.int/news-room/fact-sheets/detail/household-air-pollution-and-health
[3] Narasimhan V., Brown H., Pablos-Mendez A., et al. Responding to the global human resources crisis. The Lancet. 2004;363(9419):1469–1472. doi: 10.1016/s0140-6736(04)16108-4. [DOI] [PubMed] [Google Scholar]
[4] Naicker S., Plange-Rhule J., Tutt R. C., Eastwood J. B. Shortage of healthcare workers in developing countries. Africa, Ethnicity & Disease. 2009;19:p. 60. [PubMed] [Google Scholar]
[5] X. Wang, Y. Peng, L. Lu, Z. Lu, M. Bagheri, R.M. Summers Chestx-ray8: hospital-scale chest x-ray database and benchmarks on weakly-supervised classification and localization of common thorax diseases
Proceedings of the IEEE Conference on Computer Vision and Pattern Recognition (2017), pp. 2097-2106
[6] M. Haloi, K. R. Rajalakshmi, and P. Walia, Towards Radiologist-Level Accurate Deep Learning System for Pulmonary Screening, 2018, arXiv preprint arXiv:1807.03120.
[7] M.A. Al-masni, M.A. Al-antari, J.-M. Park, G. Gi, T.-Y. Kim, P. Rivera, et al. Simultaneous detection and classification of breast masses in digital mammograms via a deep learning YOLO-based CAD system Comput Methods Programs Biomed, 157 (2018), pp. 85-94
[8] Y. LeCun, Y. Bengio Convolutional networks for images, speech, and time series Handb. Brain Theory Neural Netw., 3361 (1995), p. 1995
[9] Wenderott, Katharina, Jim Krups, Fiona Zaruchas, and Matthias Weigl. "Effects of Artificial Intelligence Implementation on Efficiency in Medical Imaging—A Systematic Literature Review and Meta-analysis." Npj Digital Medicine 7, no. 1 (2024): 1-16. Accessed December 30, 2024. https://doi.org/10.1038/s41746-024-01248-9.
[10] Karpurapu, Shanthi, Sravanthy Myneni, Unnati Nettur, Likhit Sagar Gajja, Dave Burke, Tom Stiehm, and Jeffery Payne. "Comprehensive Evaluation and Insights into the Use of Large Language Models in the Automation of Behavior-Driven Development Acceptance Test Formulation." IEEE Access (2024).
[11] Nettur, Suresh Babu, Shanthi Karpurapu, Unnati Nettur, and Likhit Sagar Gajja. "Cypress Copilot: Development of an AI Assistant for Boosting Productivity and Transforming Web Application Testing." IEEE Access (2024).
[12] Nettur, Suresh B., Shanthi Karpurapu, Unnati Nettur, Likhit S. Gajja, Sravanthy Myneni, Akhil Dusi, and Lalithya Posham. "A Hybrid Deep Learning CNN Model for Enhanced COVID-19 Detection from Computed Tomography (CT) Scan Images." ArXiv, (2025). Accessed January 31, 2025. https://arxiv.org/abs/2501.17160.
[13] Nettur, Suresh B., Shanthi Karpurapu, Unnati Nettur, Likhit S. Gajja, Sravanthy Myneni, Akhil Dusi, and Lalithya Posham. "UltraLightSqueezeNet: A Deep Learning Architecture for Malaria Classification with up to 54x Fewer Trainable Parameters for Resource Constrained Devices." ArXiv, (2025). Accessed January 31, 2025. https://arxiv.org/abs/2501.14172.
[14] https://www.kaggle.com/datasets/paultimothymooney/chest-xray-pneumonia
[15] D.S. Kermany, M. Goldbaum, W. Cai, C.C. Valentim, H. Liang, S.L. Baxter, A. McKeown, G. Yang, X. Wu, F. Yan et al., Identifying medical diagnoses and treatable diseases by image-based deep learning. Cell 172(5), 1122–1131 (2018)
[16] E. Ayan and H. M. Ünver, "Diagnosis of Pneumonia from Chest X-Ray Images Using Deep Learning," *2019 Scientific Meeting on Electrical-Electronics & Biomedical Engineering and Computer Science (EBBT)*, Istanbul, Turkey, 2019, pp. 1-5, doi: 10.1109/EBBT.2019.8741582.
[17] Thakur, S., Goplani, Y., Arora, S., Upadhyay, R., Sharma, G. (2021). Chest X-Ray Images Based Automated Detection of Pneumonia Using Transfer Learning and CNN. In: Bansal, P., Tushir, M., Balas, V., Srivastava, R. (eds) Proceedings of International Conference on Artificial Intelligence and Applications. Advances in Intelligent Systems and Computing, vol 1164. Springer, Singapore. https://doi.org/10.1007/978-981-15-4992-2_31.
[18] Jain, Rachna, Preeti Nagrath, Gaurav Kataria, V. Sirish Kaushik, and D. Jude Hemanth. "Pneumonia Detection in Chest X-ray Images Using Convolutional Neural Networks and Transfer Learning." Measurement 165, (2020): 108046. Accessed December 11, 2024. https://doi.org/10.1016/j.measurement.2020.108046.
[19] Chhikara, Prateek & Singh, Prabhjot & Gupta, Prakhar & Bhatia, Tarunpreet. (2019). Deep Convolutional Neural Network with Transfer Learning for Detecting Pneumonia on Chest X-Rays. 10.1007/978-981-15-0339-9_13.
[20] Sara Lee Kit Yee and Wong Jee Keen Raymond. 2020. Pneumonia Diagnosis Using Chest X-ray Images and Machine Learning. In Proceedings of the 2020 10th International Conference on Biomedical Engineering and Technology (ICBET '20). Association for Computing Machinery, New York, NY, USA, 101–105. https://doi.org/10.1145/3397391.3397412
[21] El Asnaoui, K., Chawki, Y., Idri, A. (2021). Automated Methods for Detection and Classification Pneumonia Based on X-Ray Images Using Deep Learning. In: Maleh, Y., Baddi, Y., Alazab, M., Tawalbeh, L., Romdhani, I. (eds) Artificial Intelligence and Blockchain for Future Cybersecurity Applications. Studies in Big Data, vol 90. Springer, Cham. https://doi.org/10.1007/978-3-030-74575-2_14
[22] Knok, Zeljko & Pap, Klaudio & Hrncic, Marko. (2019). Implementation of intelligent model for pneumonia detection. Tehnički glasnik. 13. 315-322. 10.31803/tg-20191023102807.

[23] Liang, Gaobo, and Lixin Zheng. "A Transfer Learning Method with Deep Residual Network for Pediatric Pneumonia Diagnosis." Computer Methods and Programs in Biomedicine 187, (2020): 104964. Accessed December 11, 2024. https://doi.org/10.1016/j.cmpb.2019.06.023.

[24] Stephen O, Sain M, Maduh UJ, Jeong DU. An Efficient Deep Learning Approach to Pneumonia Classification in Healthcare. J Healthc Eng. 2019 Mar 27;2019:4180949. doi: 10.1155/2019/4180949. PMID: 31049186; PMCID: PMC6458916.

[25] Raheel Siddiqi. 2019. Automated Pneumonia Diagnosis using a Customized Sequential Convolutional Neural Network. In Proceedings of the 2019 3rd International Conference on Deep Learning Technologies (ICDLT '19). Association for Computing Machinery, New York, NY, USA, 64–70. https://doi.org/10.1145/3342999.3343001.

[26] H. Omar, A.B.-P. Book, and undefined 2019, Detection of pneumonia from X-ray images using convolutional neural network, *researchgate.net*, Accessed: Apr. 14, 2023. [Online]. Available: https://www.researchgate.net/profile/Ozge-Doguc-2/publication/344164531_Using_Dimensionality_Reduction_Techniques_to_Determine_Student_Success_Factors/links/5f577c4b92851c250b9d53c1/Using-Dimensionality-Reduction-Techniques-to-Determine-Student-Success-Factors.pdf#page=186.

[27] H. Wu, P. Xie, H. Zhang, D. Li, M. Cheng Predict pneumonia with chest X-ray images based on convolutional deep neural learning networks J. Intell. Fuzzy Syst., vol. 39 (2020), pp. 2893-2907, 10.3233/JIFS-191438

[28] S. Rajaraman, S. Candemir, G. Thoma, S. Antani Visualizing and explaining deep learning predictions for pneumonia detection in pediatric chest radiographs Proc.SPIE (Mar. 2019), 10.1117/12.2512752

[29] S. Aich, H.-C. Kim, S. Chakraborty, J. Seong Sim, and H.-C. Kim 출처, Detection of Pneumonia from Chest X-Rays using a Convolutional Neural Network Architecture. Detection of Pneumonia from Chest X-Rays using a Convolutional Neural Network Architecture. 저자 (Authors)." [Online]. Available: 〈http://www.dbpia.co.kr/journal/articleDetail?nodeId=NODE08747418〉

[30] Md.R. Hasan, S.M.A. Ullah, Md.E. Karim, 2023, An Effective Framework for Identifying Pneumonia in Healthcare Using a Convolutional Neural Network, 2023 International Conference on Electrical, Computer and Communication Engineering (ECCE), pp. 1–6, Feb. 2023, doi: 10.1109/ECCE57851.2023.10101548.

[31] Toğaçar, M., B. Ergen, Z. Cömert, and F. Özyurt. "A Deep Feature Learning Model for Pneumonia Detection Applying a Combination of MRMR Feature Selection and Machine Learning Models." IRBM 41, no. 4 (2020): 212-222. Accessed December 11, 2024. https://doi.org/10.1016/j.irbm.2019.10.006.

[32] Chouhan, Vikash, Sanjay Kumar Singh, Aditya Khamparia, Deepak Gupta, Prayag Tiwari, Catarina Moreira, Robertas Damaševičius, and Victor Hugo C. de Albuquerque. 2020. "A Novel Transfer Learning Based Approach for Pneumonia Detection in Chest X-ray Images" *Applied Sciences* 10, no. 2: 559. https://doi.org/10.3390/app10020559.

[33] Hashmi, Mohammad F., Satyarth Katiyar, Avinash G. Keskar, Neeraj D. Bokde, and Zong W. Geem. "Efficient Pneumonia Detection in Chest Xray Images Using Deep Transfer Learning." Diagnostics 10, no. 6 (2020): 417. Accessed December 11, 2024. https://doi.org/10.3390/diagnostics10060417.

[34] El Asnaoui, K. Design ensemble deep learning model for pneumonia disease classification. Int J Multimed Info Retr 10, 55–68 (2021). https://doi.org/10.1007/s13735-021-00204-7

[35] I. Alonso, L. Riazuelo and A. C. Murillo, "MiniNet: An efficient semantic segmentation ConvNet for real-time robotic applications", IEEE Trans. Robot., vol. 36, no. 4, pp. 1340-1347, Aug. 2020.

[36] A. G. Howard et al., "MobileNets: Efficient convolutional neural networks for mobile vision applications", arXiv:1704.04861, 2017.

[37] T. H. Mosbech, K. Pilgaard, A. Vaag and R. Larsen, "Automatic segmentation of abdominal adipose tissue in MRI", Proc. Scand. Conf. Image Anal., pp. 501-511, 2011.

[38] J. Zhang, Y. Xie, P. Zhang, H. Chen, Y. Xia and C. Shen, "Light-weight hybrid convolutional network for liver tumor segmentation", Proc. 28th Int. Joint Conf. Artif. Intell., pp. 4271-4277, Aug. 2019.

[39] M. Sandler, A. Howard, M. Zhu, A. Zhmoginov and L.-C. Chen, "MobileNetV2: Inverted residuals and linear bottlenecks", Proc. IEEE/CVF Conf. Comput. Vis. Pattern Recognit., pp. 4510-4520, Jun. 2018.

[40] Dong, Ke, Chengjie Zhou, Yihan Ruan, and Yuzhi Li. "MobileNetV2 model for image classification." In 2020 2nd International Conference on Information Technology and Computer Application (ITCA), pp. 476-480. IEEE, 2020.

[41] Gulzar, Yonis. "Fruit image classification model based on MobileNetV2 with deep transfer learning technique." Sustainability 15, no. 3 (2023): 1906.

[42] Xiang, Qian, Xiaodan Wang, Rui Li, Guoling Zhang, Jie Lai, and Qingshuang Hu. "Fruit image classification based on Mobilenetv2 with transfer learning technique." In Proceedings of the 3rd international conference on computer science and application engineering, pp. 1-7. 2019.

[43] Liu, Jun, and Xuewei Wang. "Early recognition of tomato gray leaf spot disease based on MobileNetv2-YOLOv3 model." Plant Methods 16 (2020): 1-16.

[44] Sanjaya, Samuel Ady, and Suryo Adi Rakhmawan. "Face mask detection using MobileNetV2 in the era of COVID-19 pandemic." In 2020 International Conference on Data Analytics for Business and Industry: Way Towards a Sustainable Economy (ICDABI), pp. 1-5. IEEE, 2020.

[45] Nagrath, Preeti, Rachna Jain, Agam Madan, Rohan Arora, Piyush Kataria, and Jude Hemanth. "SSDMNV2: A real time DNN-based face mask detection system using single shot multibox detector and MobileNetV2." Sustainable cities and society 66 (2021): 102692.

[46] A. M. K. Izzaty, T. W. Cenggoro, G. N. Elwirehardja and B. Pardamean, "Multiclass classification of histology on colorectal cancer using deep learning", *Commun. Math. Biol. Neurosci.*, pp. 1-19, Jul. 2022.

[47] M. Afif, R. Ayachi, Y. Said and M. Atri, "Deep learning based application for indoor scene recognition", *Neural Process. Lett.*, vol. 51, no. 3, pp. 2827-2837, Mar. 2020.

[48] H. A. Shah, F. Saeed, S. Yun, J. -H. Park, A. Paul and J. -M. Kang, "A Robust Approach for Brain Tumor Detection in Magnetic Resonance Images Using Finetuned EfficientNet," in *IEEE Access*, vol. 10, pp. 65426-65438, 2022, doi: 10.1109/ACCESS.2022.3184113.

[49] B. A. Tama, M. Vania, I. Kim and S. Lim, "An EfficientNet-Based Weighted Ensemble Model for Industrial Machine Malfunction Detection Using Acoustic Signals," in *IEEE Access*, vol. 10, pp. 34625-34636, 2022, doi: 10.1109/ACCESS.2022.3160179.

[50] Marques, Gonçalo, Deevyankar Agarwal, and Isabel De la Torre Díez. "Automated medical diagnosis of COVID-19 through EfficientNet convolutional neural network." Applied soft computing 96 (2020): 106691.

[51] Atila, Ümit, Murat Uçar, Kemal Akyol, and Emine Uçar. "Plant leaf disease classification using EfficientNet deep learning model." Ecological Informatics 61 (2021): 101182.

[52] Duong, Linh T., Phuong T. Nguyen, Claudio Di Sipio, and Davide Di Ruscio. "Automated fruit recognition using EfficientNet and MixNet." Computers and Electronics in Agriculture 171 (2020): 105326.

[53] Nayak, Dillip Ranjan, Neelamadhab Padhy, Pradeep Kumar Mallick, Mikhail Zymbler, and Sachin Kumar. "Brain tumor classification using dense efficient-net." Axioms 11, no. 1 (2022): 34.

[54] Wang, Jing, Liu Yang, Zhanqiang Huo, Weifeng He, and Junwei Luo. "Multi-label classification of fundus images with efficientnet." IEEE access 8 (2020): 212499-212508.

[55] Coccomini, Davide Alessandro, Nicola Messina, Claudio Gennaro, and Fabrizio Falchi. "Combining efficientnet and vision transformers for video deepfake detection." In International conference on image analysis and processing, pp. 219-229. Cham: Springer International Publishing, 2022.

[56] M. Tan and Q. V. Le, "EfficientNet: Rethinking model scaling for convolutional neural networks", *Proc. Int. Conf. Mach. Learn.*, pp. 6105-6114, 2019.

[57] B. Zoph, V. Vasudevan, J. Shlens and Q. V. Le, "Learning transferable architectures for scalable image recognition", *Proc. IEEE/CVF Conf. Comput. Vis. Pattern Recognit.*, pp. 8697-8710, Jun. 2018.

[58] Adedoja, Adedamola O., Pius A. Owolawi, Temitope Mapayi, and Chunling Tu. "Intelligent mobile plant disease diagnostic system using NASNet-mobile deep learning." IAENG International Journal of Computer Science 49, no. 1 (2022): 216-231.

[59] Ahsan, Md Manjurul, Kishor Datta Gupta, Mohammad Maminur Islam, Sajib Sen, Md Lutfar Rahman, and Mohammad Shakhawat Hossain. "Covid-19 symptoms detection based on nasnetmobile with explainable ai using various imaging modalities." Machine Learning and Knowledge Extraction 2, no. 4 (2020): 490-504.

[60] Saxen, Frerk, Philipp Werner, Sebastian Handrich, Ehsan Othman, Laslo Dinges, and Ayoub Al-Hamadi. "Face attribute detection with mobilenetv2 and nasnet-mobile." In 2019 11th international symposium on image and signal processing and analysis (ISPA), pp. 176-180. IEEE, 2019.

[61] P. K. Das and S. Meher, "Transfer Learning-Based Automatic Detection of Acute Lymphocytic Leukemia," *2021 National Conference on Communications (NCC)*, Kanpur, India, 2021, pp. 1-6, doi: 10.1109/NCC52529.2021.9530010.

[62] S. Verma, M. A. Razzaque, U. Sangtongdee, C. Arpnikanondt, B. Tassaneetrithep and A. Hossain, "Digital Diagnosis of Hand, Foot, and Mouth Disease Using Hybrid Deep Neural Networks," in *IEEE Access*, vol. 9, pp. 143481-143494, 2021, doi: 10.1109/ACCESS.2021.3120199.

[63] Lin, Min, Qiang Chen, and Shuicheng Yan. "Network In Network." ArXiv, (2013). Accessed December 4, 2024. https://arxiv.org/abs/1312.4400.

[64] Srivastava, Nitish, Geoffrey Hinton, Alex Krizhevsky, Ilya Sutskever, and Ruslan Salakhutdinov. "Dropout: a simple way to prevent neural networks from overfitting." The journal of machine learning research 15, no. 1 (2014): 1929-1958.

[65] Ioffe, Sergey. "Batch normalization: Accelerating deep network training by reducing internal covariate shift." arXiv preprint arXiv:1502.03167 (2015).

**SURESH BABU NETTUR** has received his Master of Science (M.S) degree from the Birla Institute of Technology and Science (BITS), Pilani, Rajasthan, India, and his Bachelor of Technology degree in Computer Science and Engineering from Nagarjuna University, Guntur, India. With over two decades of expertise, he has established himself as a thought leader in artificial intelligence, deep learning, and machine learning solutions. His work spans the development of scalable and intelligent systems across industries such as healthcare, finance, telecom, and manufacturing. Suresh has been at the forefront of integrating AI-driven innovations into real-world applications, leveraging cutting-edge technologies such as OpenAI models, GitHub Copilot, and custom deep learning architectures to deliver transformative solutions. His contributions include designing and implementing advanced machine learning models, optimizing deep learning architectures for resource-constrained environments, and integrating AI solutions into software development and testing pipelines. With significant experience in cross-functional team leadership and managing onsite-offshore collaboration models, he has successfully delivered AI-powered applications across cloud platforms like AWS.

Suresh is passionate about applying AI to solve complex problems in healthcare and finance, including predictive analytics, automation, and intelligent decision-making systems. He is proficient in Agile methodologies, Test-Driven Development (TDD), and service-oriented architectures (SOA), ensuring seamless integration of AI and machine learning into software systems. As an advocate for innovative AI applications, Suresh is committed to advancing the field through sustainable and impactful solutions that redefine industry standards and improve quality of life.

**SHANTHI KARPURAPU** received the Bachelor of Technology degree in chemical engineering from Osmania University, Hyderabad, India and the Masters technology degree in chemical engineering from Institute of Chemical Technology, Mumbai, India.
She has over a decade of experience leading, designing, and developing test automation solutions for various platforms across healthcare, banking, and manufacturing industries using Agile and Waterfall methodologies. She is experienced in building reusable and extendable automation frameworks for web applications, REST, SOAP, and microservices. She is a strong follower of the shift-left testing approach, a certified AWS Cloud practitioner, and a machine learning specialist. She is passionate about utilizing AI-related technologies in software testing and the healthcare industry.

**UNNATI NETTUR** currently pursuing an undergraduate degree in Computer Science at Virginia Tech, Blacksburg, VA, USA. She possesses an avid curiosity about the constantly evolving field of technology and software development, with a particular interest in Artificial Intelligence. She is passionate about gaining experience in building innovative and creative solutions for current issues in the field of software engineering.

**LIKHIT SAGAR GAJJA** pursuing a Computer Science Bachelor's degree at BML Munjal University, Haryana, INDIA. He is evident in showing his passion for the dynamic field of technology and software development. His specific interests include Artificial Intelligence, Prompt Engineering, and Game Designing technologies, highlighting his dedication to obtaining hands-on experience and developing innovative solutions for real-time issues in software engineering.

**SRAVANTHY MYNENI** earned master of science in information technology and management from Illinois institute of technology, Chicago, Illinois in 2017 and Bachelor's degree in computer science in 2013. She is currently working as an engineer focused on data engineering and analysis. She has 8+ years of experience in designing, building and deploying data centric solutions using Agile and Waterfall methodologies. She is enthusiastic about data analysis, data engineering and AI application to provide solutions for real world problems.

**AKHIL DUSI** currently pursuing Masters of Information Sciences at University of Indiana Tech, Indiana, USA. He is a passionate researcher and developer with a diverse background in software development, cybersecurity, and emerging technologies. He has a proven ability to deliver innovative and practical solutions. His work includes developing web and mobile-based applications, conducting vulnerability assessments and penetration testing, and leveraging cloud platforms for efficient infrastructure management. He is certified in cybersecurity and machine learning, reflecting a strong commitment to continuous learning and staying at the forefront of technological advancements. His research interests focus on artificial intelligence, IoT, and secure system design, with a vision to drive impactful innovations.

**LALITHYA POSHAM** is an MBBS graduate from Nanjing Medical University. She has a strong passion for advancing clinical research and improving patient outcomes. With a solid foundation in medical education, she is particularly interested in exploring innovative diagnostic approaches and aims to integrate clinical expertise with research to drive improvements in healthcare systems.